\documentclass[a4paper, conference]{IEEEtran}
\IEEEoverridecommandlockouts
\usepackage{cite}
\usepackage{amsmath,amssymb,amsfonts}
\usepackage{algorithmic}
\usepackage{graphicx}
\usepackage{textcomp}
\usepackage{xcolor}
\usepackage[ruled,vlined]{algorithm2e}
\usepackage{csvsimple}
\usepackage[utf8]{inputenc}

\DeclareUnicodeCharacter{202F}{\,}

\def\BibTeX{{\rm B\kern-.05em{\sc i\kern-.025em b}\kern-.08em
    T\kern-.1667em\lower.7ex\hbox{E}\kern-.125emX}}

\makeatletter

\def\ps@IEEEtitlepagestyle{%
  \def\@oddfoot{\mycopyrightnotice}%
  \def\@evenfoot{}%
}
\def\mycopyrightnotice{%
  \gdef\mycopyrightnotice{}
}

\begin{document}

\title{Enhancement of Distribution System State Estimation Using Pruned Physics-Aware Neural Networks\\

\thanks{The authors would like to acknowledge the financial support for this work from the Enabling flexibility for future distribution grid project FlexiGrid (EC funding number: 864048). This work was authored in part by the National Renewable Energy Laboratory (NREL), operated by Alliance for Sustainable Energy, LLC, for the U.S. Department of Energy (DOE) under Contract No. DE-AC36-08GO28308. The work of A. S. Zamzam was supported by the Laboratory Directed Research and Development (LDRD) Program at NREL. The views expressed in the article do not necessarily represent the views of the DOE or the U.S. Government. The U.S. Government retains and the publisher, by accepting the article for publication, acknowledges that the U.S. Government retains a nonexclusive, paid-up, irrevocable, worldwide license to publish or reproduce the published form of this work, or allow others to do so, for U.S. Government purposes.}}

\author{\IEEEauthorblockN{Minh-Quan Tran}
\IEEEauthorblockA{\textit{Department of Electrical Engineering} \\
\textit{Eindhoven University of Technology}\\
Eindhoven, The Netherlands\\
m.q.tran@tue.nl\\}

\and
\IEEEauthorblockN{Ahmed S. Zamzam}
\IEEEauthorblockA{\textit{National Renewable Energy Laboratory} \\
Golden, CO, United States\\
Ahmed.Zamzam@nrel.gov}

\and
\IEEEauthorblockN{Phuong H. Nguyen}
\IEEEauthorblockA{\textit{Department of Electrical Engineering} \\
Eindhoven University of Technology\\
Eindhoven, The Netherlands\\
p.nguyen.hong@tue.nl}

}

\maketitle

\begin{abstract}
Realizing complete observability in the three-phase distribution system remains a challenge that hinders the implementation of classic state estimation algorithms. In this paper, a new method, called the pruned physics-aware neural network (P2N2), is developed to improve the voltage estimation accuracy in the distribution system. The method relies on the physical grid topology, which is used to design the connections between different hidden layers of a neural network model. To verify the proposed method, a numerical simulation based on one-year smart meter data of load consumptions for three-phase power flow is developed to generate the measurement and voltage state data. The IEEE 123-node system is selected as the test network to benchmark the proposed algorithm against the classic weighted least squares (WLS). Numerical results show that P2N2 outperforms WLS in terms of data redundancy and estimation accuracy.
\end{abstract}


\section{Introduction}
State estimation is an important function for grid monitoring and control. The traditional weighted least squares (WLS) is often used to estimate the system state (e.g., voltage magnitude, voltage angle). Different from transmission systems, distribution systems are nominally unobservable \cite{Dehghanpour2019a,Tran2020}. Caused by the scarcity of measurement devices, WLS is no longer applicable in a more extensive distribution system because the singularity of the gain matrix hinders the solvability for the state variables \cite{Primadianto2017}.

A practical solution of the unobservable grid is to use pseudo measurements, which are forecasted from historical data or calculated by interpolating observed measurements data. In distribution systems, pseudo measurements can be obtained from smart meter data, distributed energy resource generation based on the forecasting model of photovoltaic (PV) irradiance or wind speed. In \cite{Dehghanpour2019d}, a game theoretic-based data-driven technique is studied with the purpose of generating pseudo measurements in distribution system state estimation (DSSE). A neural network model is developed in \cite{Biserica2012} to form unbiased load distributions to improve Volt-Var control in distribution grid. The parallel machine learning model is developed to learn load patterns and then to generate accurate active power pseudo measurements. For the same purpose, in \cite{Gahrooei2018d}, a frequency-based clustering algorithm is implemented, which determines the load patterns and estimates the daily energy consumption. On the other hand, a probabilistic data-driven method is used to generate time-series pseudo measurements for unmeasured PV systems \cite{Yuan2019f}.

Besides exploiting pseudo measurements from abundant data to improve the grid monitoring, distribution system operators (DSOs) will benefit from methods that can predict the system state with limited sensing. An estimation method with a combination of forecasting and the state estimation model was proposed in \cite{Dobbe2020a,Zamzam2018,Zhang2018,Mestav2019a}. These methods proposed data-driven models, which rely on minimum mean squares estimation and Bayesian estimation. The advantage is that these methods do not require observability or redundant measurements. Recently, the authors in \cite{Mestav2019a} proposed a deep learning-based Bayesian state estimation approach for unobservable distribution grids. The data-driven techniques present a very promising solution to improve grid observability in distribution systems. Motivated by these approaches, we propose a data-driven state estimation with limited sensing to solve the problem DSOs are facing. In \cite{Zamzam2020a}, a method called the physics-aware neural network (PAWNN) model was proposed. The idea is to embed the physical connection of the distribution system into the neural network model; however, the connection between consecutive layers in the model is kept the same, which leads to possible unnecessary connections. To this end, this paper proposes the pruned physics-aware neural network (P2N2).

\begin{figure}[t]
\centerline{\includegraphics[width=0.85\linewidth]{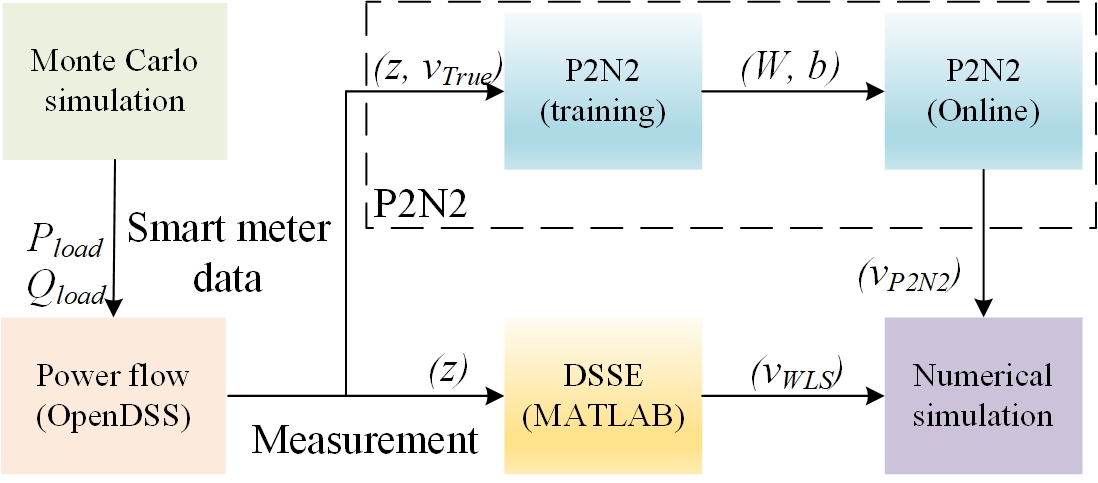}}
\caption{The methodology.}
\label{fig1}
\vspace{-5mm}
\end{figure}

A graphic summary of the proposed approach is shown in Fig. 1. First, Monte Carlo simulations are set up with 35,040 data points from smart meters (one-year collected data). Then, the load consumption data are fed into the power flow simulation. The voltage magnitude and measurement data are then used as the input for the DSSE and the P2N2 model. Three different simulation cases are carried out with the modified IEEE 123-node test system. The simulation results show the effectiveness of the P2N2 approach.

\section{Distribution System State Estimation}
In this section, the three-phase estimator is presented in Part A, which is a rectangular voltage-based DSSE. In Part B, different types of measurements in the distribution system are discussed in detail.
\subsection{Three-Phase Estimator}
The state estimation is a well-known method that aims to estimate the system's state variables from measurements based on the mathematical relations between system states and measurement points. The synthesizing function can be written as follows:
\begin{equation}
z=h(x)+\ e\label{eq1}
\end{equation}
where $z$ is a measurement vector obtained from grid measurements and pseudo measurements; $h(x)$ is a vector function from state variables, $x$, to measurements, $z$; $e$ is a measurement noise vector that is assumed to be independent zero-mean Gaussian variables. The covariance of the measurement noise is denoted by $R$. In general, the objective of the WLS method is minimizing the sum of the square of the residuals:
\begin{equation}
J\left(x\right)=\left[z-h(x)\right]^TR^{-1}[z-h(x)].\label{eq2}
\end{equation}
The Gauss-Newton method is usually used to obtain the solution. The iterative process stops when the number of iterations is higher than the limited value or when the changes in residual are less than the limited tolerance (usually as 10E-5 or 10E-7).
\begin{equation}
x^k=x^{k-1}+{G\left(x^k\right)}^{-1}{H\left(x^k\right)}^TR^{-1}\left[z-h\left(x\right)\right]\label{eq3}
\end{equation}
where $H(x)$ and $G(x)$ are:
\begin{equation}
H\left(x\right)=\frac{\partial h(x)}{\partial x}\label{eq4}
\end{equation}
\begin{equation}
G\left(x^k\right)={H\left(x^k\right)}^TR^{-1}{H\left(x^k\right)}\label{eq5}
\end{equation}
Different from transmission systems, distribution networks are highly unbalanced systems. This leads to singularity of the gain matrix, $G\left(x^k\right)$, and hence the single-phase state estimation model used for transmission state estimation is often not applicable for DSSE. In this work, the three-phase state estimator from \cite{Muscas2014} is used, which is based on the rectangular voltage. The state variables of the network are represented by three-phase rectangular form (i.e., the real part and imaginary part) at every node.

\subsection{The Used Measurements}

Because distribution systems are highly unobservable, the application of the WLS algorithm needs additional pseudo measurements to remedy the low-observability issue. The measurements used in this work are:

\begin{enumerate}
 \item Phasor measurement units (PMUs): this is the three-phase synchronized measurement. Normally, it is located near the step-down transformer, which is used to measure the voltage phasor at the node and current phasor of the connected branches. With PMUs, the maximum error is 1\% for the magnitude and 10E-2 rad for the phase angle.
\item Smart meters (SMs): these measurements are installed at the household (customer measurements). The power consumption of customers is obtained normally every 15 minutes. With SMs, the maximum error is 2\% for power measurement.
\item Pseudo measurements: the historical data are used at the buses where no measurement device is installed, which can be calculated or forecasted from existing information. The three-phase active and reactive power can be obtained as pseudo measurements. The maximum error of the pseudo measurement could be up to 50\% for active and reactive power absorbed from loads.
\item Zero injection buses: the buses without any loads or generators connected are considered zero injection buses. The active and reactive power injection measured at these buses is zero, with maximum error equal to 0.001\%.
\end{enumerate}

\section{Pruned Physics-Aware Neural Network}
In this section, the proposed method of P2N2 is discussed in detail. The background of the partitioning of the DSSE based on the PMU location is explained in Part A. An example of a 6-bus system is presented. We show the way we design the P2N2 based on the physical distribution grid. Then, in Part B, we present the model validation.

\subsection{Layer Design-Based Physical Model}\label{AA}

\begin{figure}[b]
\vspace{-3mm}
\centerline{\includegraphics[width=0.55\linewidth]{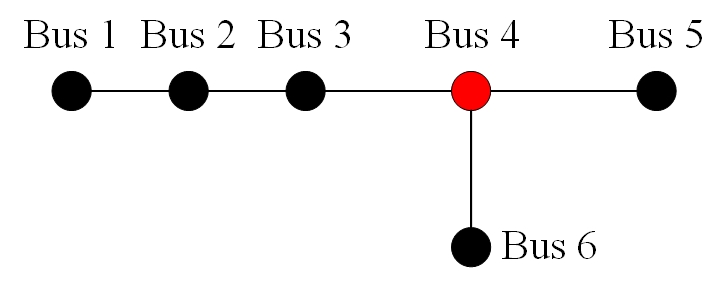}}
\caption{Example of 6-bus system with PMU located at Bus 4.}
\label{fig2}
\vspace{-3mm}
\end{figure}
As mentioned earlier, the PMU is a three-phase synchronized measurement of the real-time measured value with very high accuracy. Considering this advantage of the PMU measurement, the estimated voltage at a specified bus does not require the information of all available measurements in the network. This means that with an accurate measurement at a specified bus, other measurements behind this bus can be neglected. This separability property was proven in \cite{Zamzam2020a}. As an example, Fig. 2 shows a simple 6-bus system with a PMU installed at Bus 4. Applying the concept of vertex-cut, the system can be divided into three different partitions, as shown in Fig. 3.

\begin{enumerate}
 \item Partition 1 in Fig. 3 (a): buses 1, 2, 3 and 4.
\item Partition 2 in Fig. 3 (b): buses 4 and 5.
\item Partition 3 in Fig. 3 (c): buses 4 and 6.
\end{enumerate}

\begin{figure}[t]
\centerline{\includegraphics[width=0.55\linewidth]{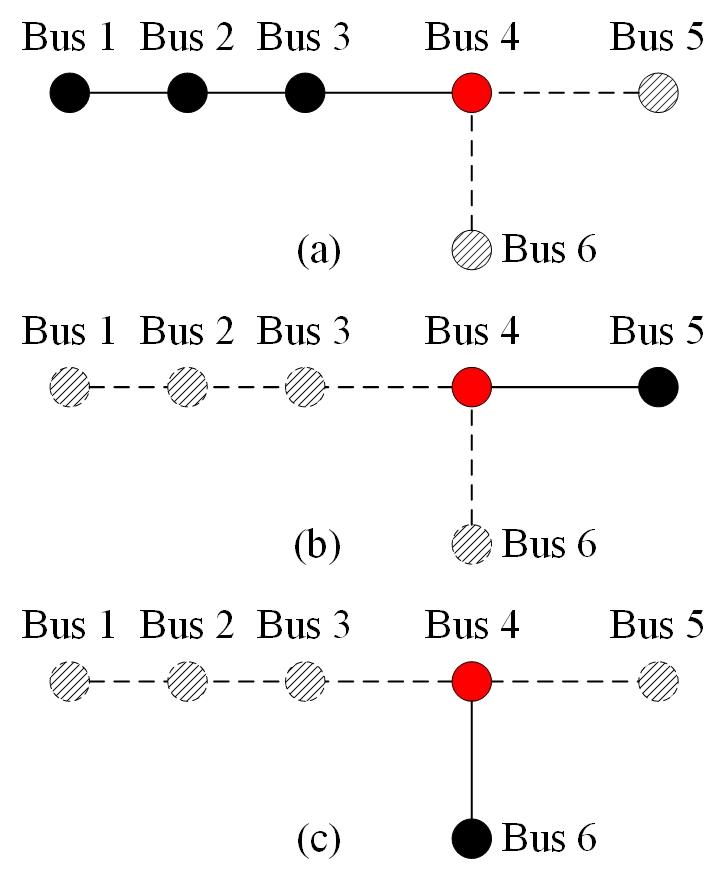}}
\caption{Vertex-cut partitioning example with PMU located at Bus 4. (a) Partition 1 with buses 1, 2, 3, and 4. (b) Partition 2 with buses 4 and 5. Partition 3 with buses 4 and 6.}
\label{fig3}
\vspace{-5mm}
\end{figure}

\begin{figure}[t]
\vspace{2mm}
\centerline{\includegraphics[width=0.8\linewidth]{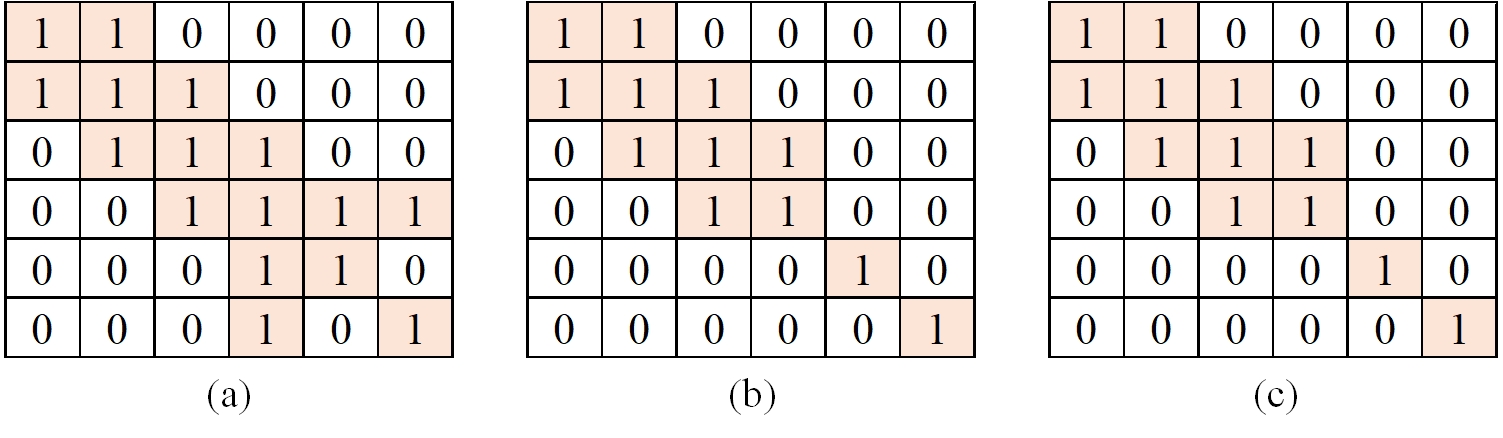}}
\caption{The designed connection between layers for 6-bus system. (a) the connection between the input layer and Layer 2. (b) the connection between Layers 2 and 3. (c) the connection between Layer 3 and the output layer.}
\label{fig4}
\vspace{-5mm}
\end{figure}

The PAWNN model is designed with multiple layers, which are built based on the physical connections of the distribution network. The required number of layers is the maximum diameter of each partition. In this case, the number of hidden layers is 3 because the maximum diameter of all partitions is 3. Then, the connections between layers are designed based on the physical connection of the network. This is the idea behind the physics-aware technique, which prunes the connections that are not present in the physical network. Fig. 4 shows the result of the designed connections between layers for the 6-bus system. Fig. 4 (a) shows exactly the structure of the network admittance matrix of the 6-bus network.

Let the input layer of the PAWNN model be denoted by $x$, and $y$ is the output layer of PAWNN. The output vector, y, represents the voltage at the buses of the network. Let $k(i)$ denoted the intermediate output at $i$-$th$ layer. Then, we have:

\begin{equation}
k_{i+1}=\sigma_i(W_ik_i)\label{eq6}
\end{equation}
where $\sigma_i$ is a point-wise nonlinearity; and $W$ has a size of $N$x$N$ weight matrix, with $N$ the number of output $y$. The matrix $W$ is designed the same as the connection shown in Fig. 4 (a). The $(i,j)$ element in the matrix $W$ is pruned if nodes $i$ and $j$ are not connected. Then, the structure of PAWNN is shown in Fig. 5 (a); however, this structure leads to possible unnecessary connections. Partitions 2, and 3 have the same diameter of 2, meaning that we can get the voltage values of buses 5 and 6 after Layer 2. To this end, the P2N2 is proposed to reduce unnecessary connections between layers. As shown in Fig. 4 (b), the connection between Layer 2 and Layer 3---four unnecessary connections of (4,5), (4,6), (5,4), and (6,4)---are zeroed out. Similarly, the new structure of the connection between Layer 3 and the output layer is shown in Fig. 4 (c). Then, three different weight matrices are used for the P2N2 model, which is shown in Fig. 5 (b). Therefore, the output of the P2N2 can be written as:
\begin{equation}
y_i=\begin{cases}\sigma_3(W_3\sigma_2(W_2\sigma_1(W_1x+b_1)+b_2)+b_3)\text{ if $i$=1,2,3,4} \\
\sigma_2(W_2\sigma_1(W_1x+b_1)+b_2)\text{ if $i$=5,6} \end{cases}\label{eq6}
\end{equation}
where $b_1$,  $b_2$,  $b_3$ are the bias vectors of the P2N2 model.

\begin{figure}[t]
\centerline{\includegraphics[width=0.8\linewidth]{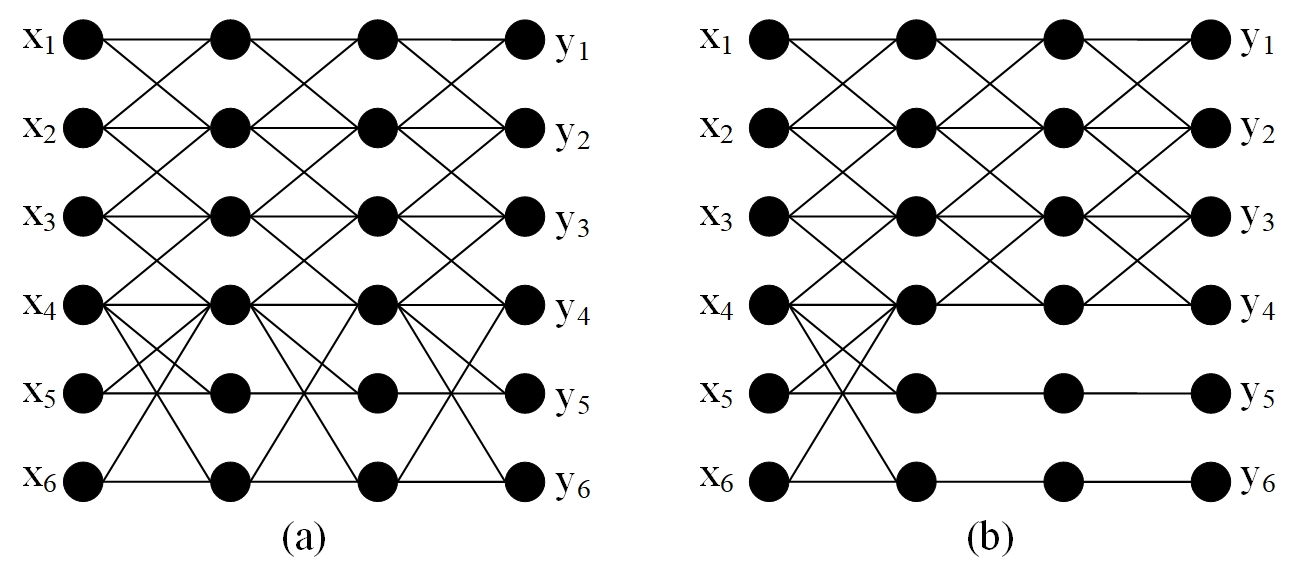}}
\caption{The graph-pruned neural network model. (a) The structure of PAWNN. (b) The designed structure of P2N2.}
\label{fig5}
\vspace{-7mm}
\end{figure}
\subsection{Model Validation}
In this work, TensorFlow \cite{Abadi2016} was used to train the model. The data were divided into 90\% training and 10\% testing. The model was trained based on the ADAM optimizer \cite{Kingma2015}, and the optimization function is formulated as follows:

\begin{equation}
\min\limits_{ \left\lbrace b_t, W_t \right\rbrace_{t=1}^T} \sum_j \Vert v^j - g_T(z^j;\left\lbrace b_t, W_t \right\rbrace_{t=1}^T)\Vert_2^2
\label{eq6}
\end{equation}where $v^j$ and $z^j$ are the true state and measurement in the $j$-$th$ training sample, respectively. $g_T$ is the $j$-$th$ mapping realized by the T-layer of the model parameterized by $\left\lbrace b_t, W_t \right\rbrace_{t=1}^T$. The network structures are imposed on the P2N2 model; hence, the number of neurons in each layer is proportional to the number of buses. Finally, we used the average estimation to calculate the accuracy of each algorithm as follows:
\begin{equation}
\nu= \frac{1}{N}\sum_{i=1}^N \Vert \hat{v}^i - v_{true}^i\Vert_2^2
\label{eq_performance}
\end{equation}
where $\hat{v}^i$ is the estimated voltage.

\section{Simulation and Results}
In this section, the test case and the simulation results are presented. First, the IEEE 123-node test system is described. Then, the methodologies explained in sections II and III are applied. Three different scenarios were carried out to assess the performance of the proposed model.

\subsection{Test Network}
In this work, the IEEE 123-node test system is used, as shown in the grid topology in Fig. 6. The IEEE 123-node system is a radial distribution grid with single-phase loads and two-phase loads; thus, the grid is a highly unbalanced network. The grid has four different voltage regulators and different voltage levels. The detailed grid parameters are available in \cite{Schneider2018}. There are four switches (13-152, 60-160, 97-197, and 18-135), which have been modified as connection buses. Further, voltage regulators are excluded in this work. Generally, these modifications are common for this kind of study \cite{Muscas2014} without affecting the generality of the study. The DSSE algorithm is built in the MATLAB environment, and the OpenDSS is used for the power flow calculation. In addition, we assumed that the system has two PMUs, one each installed at Bus 149 and Bus 60. Then, the whole system can be split into two partitions. The first area in the dashed line has one PMU, at Bus 149. By exporting the maximum diameter of possible partitions in this area, the diameter is 14. Similarly, in the second area, the diameter is 11. Then, the neural network model is designed with 14 layers.

\begin{figure}[t]
\centerline{\includegraphics[width=0.85\linewidth]{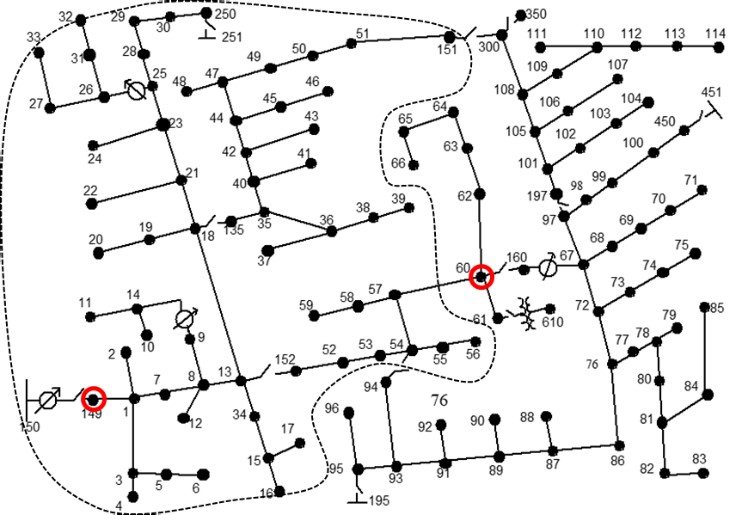}}
\caption{The IEEE 123-node test system}
\label{fig6}
\vspace{-5mm}
\end{figure}
\begin{figure}[t]
\centerline{\includegraphics[width=0.85\linewidth]{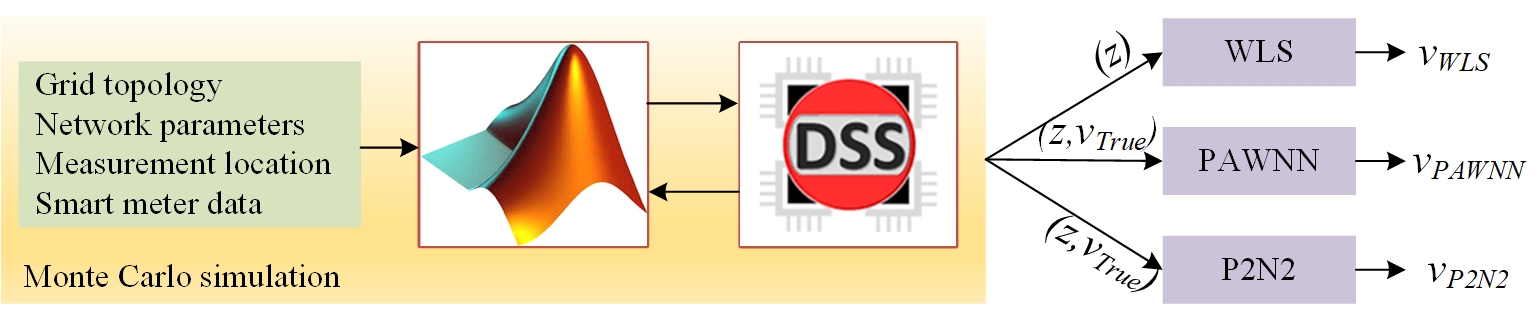}}
\caption{The simulation and model evaluation process.}
\label{fig7}
\vspace{-5mm}
\end{figure}
\subsection{Simulation Scenarios}
To perform the behavior of the estimator, one-year time-series collected data of the SM are used with 35,040 data points. Then, M= 35,040 possible operation conditions (normally, a data set of 10,000 is sufficient to ensure the quality of the results) is fed into the power flow model as the load consumption. By extracting from each power flow simulation, the measurements and the true voltage magnitude values at the buses are collected. Hence, we have 35,040 sets of measurements (z) for the WLS and 35,040 sets of measurements (z) and system states (voltage magnitude, $v_{true}$) for the P2N2. In addition, 90\% of the 35,040 sets of data is used for training, and the rest is used for testing. The process of the simulation and model evaluation is shown in Fig. 7. Then, the performance of each algorithm is calculated using (\ref{eq_performance}). This whole process is tested with three different scenarios:

\begin{figure}[t]
\centerline{\includegraphics[width=0.9\linewidth]{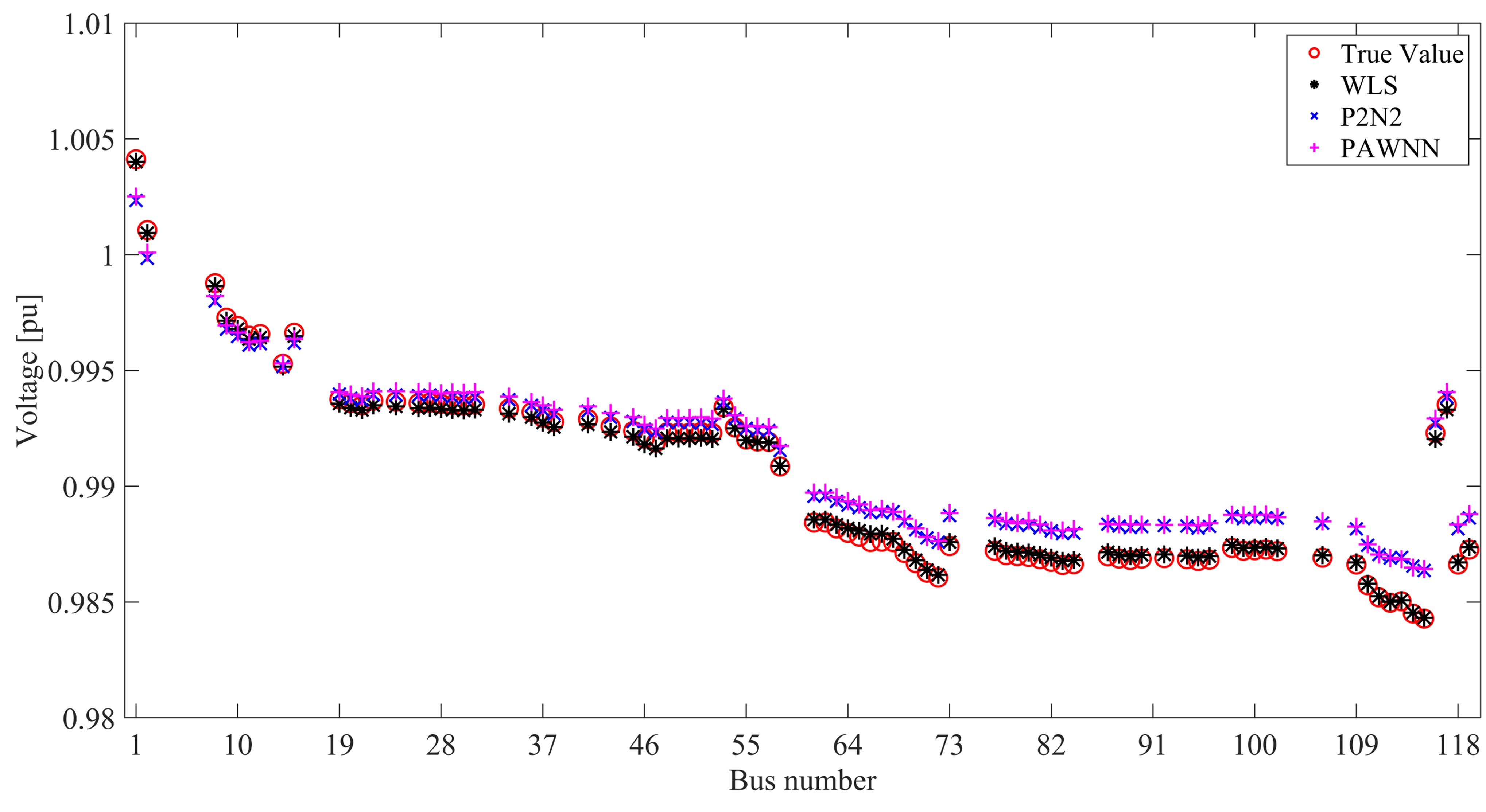}}
\caption{Estimated voltage magnitude at Phase A in Scenario 1.}
\label{fig8}
\vspace{-6mm}
\end{figure}
\begin{figure}[t]
\vspace{2mm}
\centerline{\includegraphics[width=0.9\linewidth]{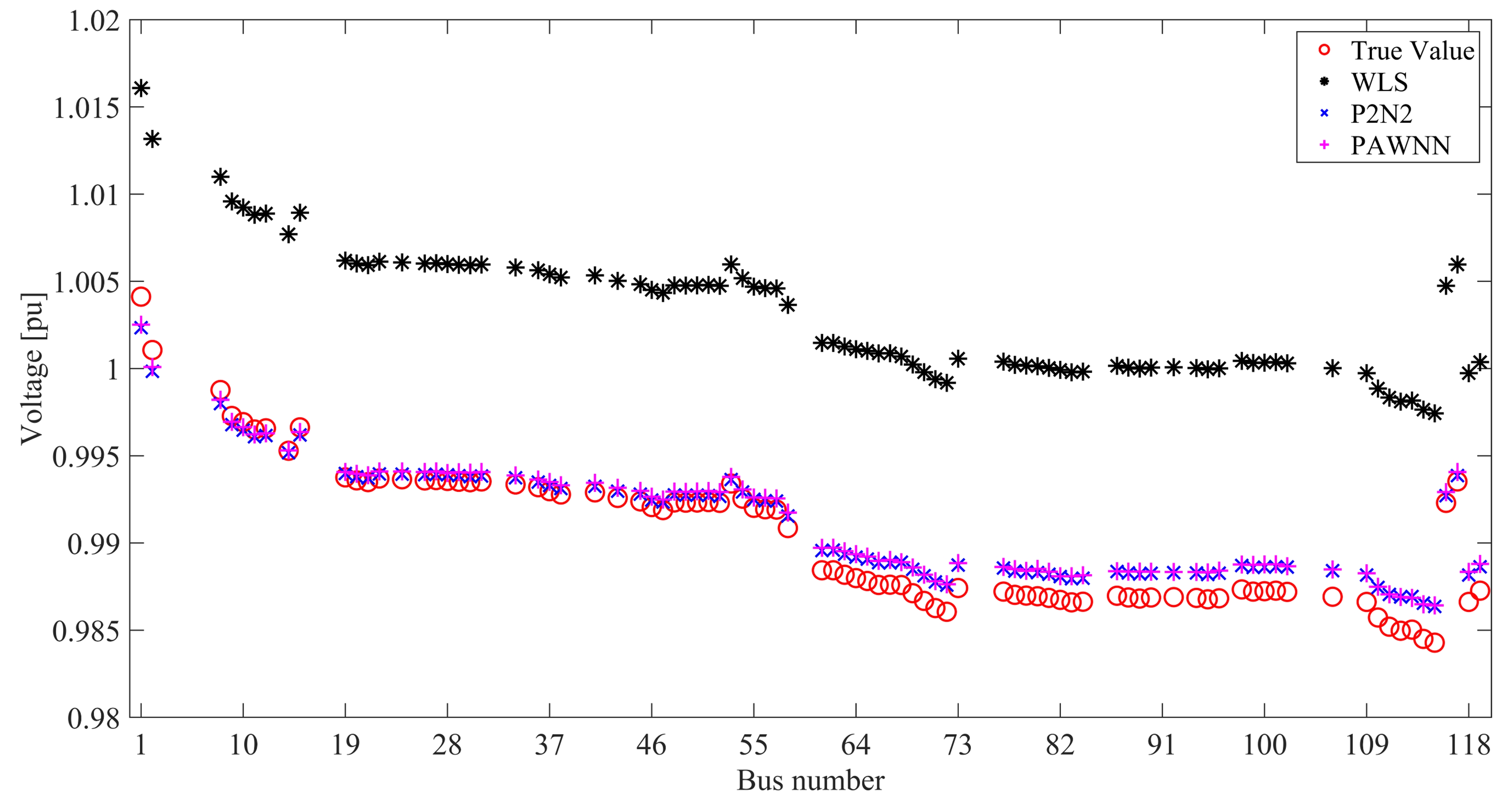}}
\caption{Estimated voltage magnitude at Phase A in Scenario 2.}
\label{fig9}
\vspace{-5mm}
\end{figure}
\begin{figure}[t]
\centerline{\includegraphics[width=0.9\linewidth]{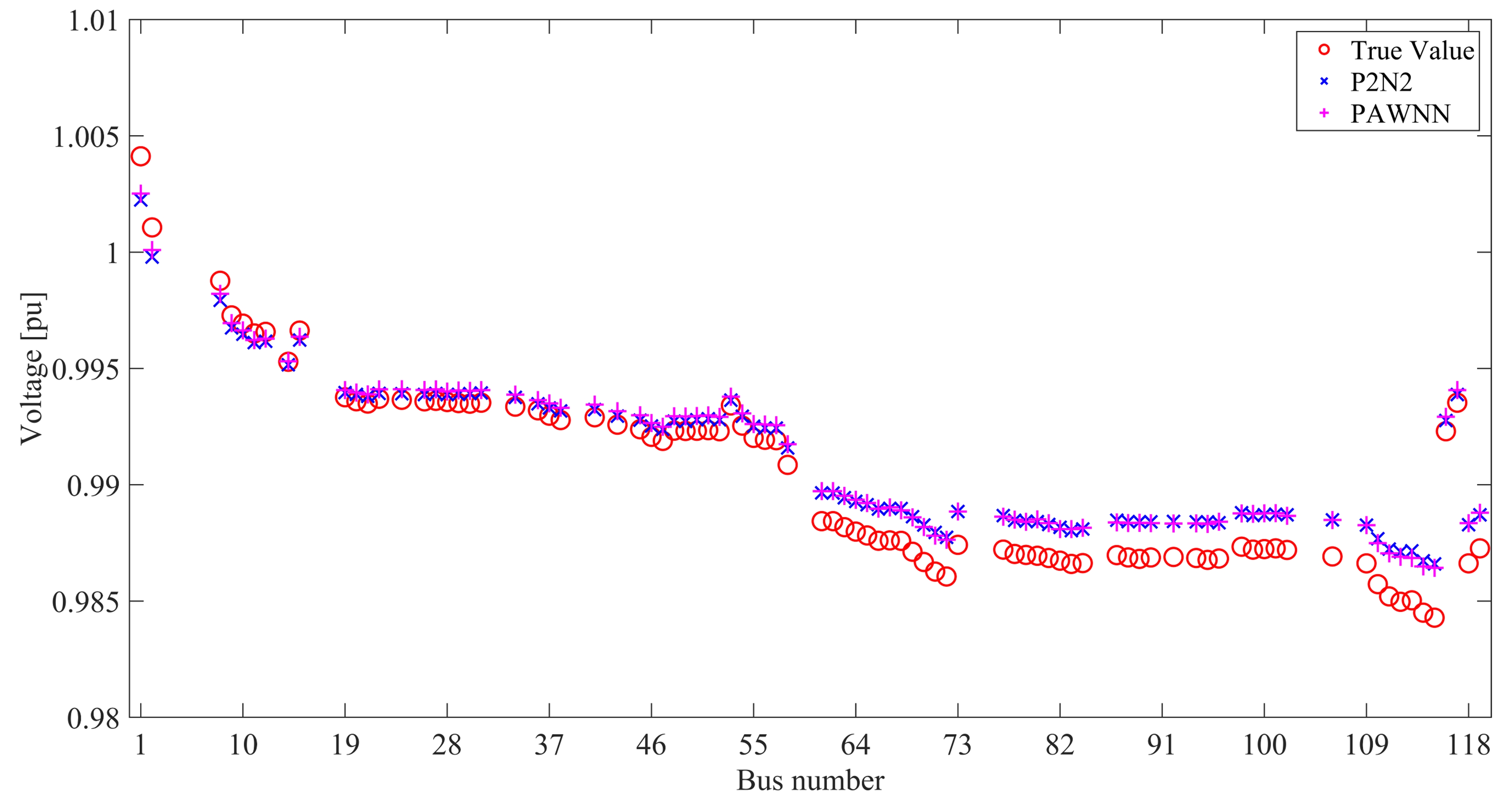}}
\caption{Estimated voltage magnitude at Phase A in Scenario 3.}
\label{fig10}
\vspace{-4mm}
\end{figure}

\begin{enumerate}
 \item The algorithms are tested with a large number of measurements.
\item We kept the same amount of measurements, and increased the error of the pseudo measurements from 30\% to 50\%.
\item Limited measurements are used for this scenario, i.e., 14 pseudo measurements are removed.
\end{enumerate}

In the first scenario, the network has 2 voltage measurements and 2 current injection measurements at Bus 149 and Bus 60. Further, 118 pseudo measurements are used, which consist of 85 load power measurements and 33 zero injection measurements. As an example, we present only the voltage magnitudes at Phase A of all the buses. Fig. 8 depicts the estimated voltage magnitudes in the first scenario. The results show the robustness of the WLS in case of redundant measurements. However, the PAWNN and P2N2 also show the high accuracy of the estimated voltage magnitudes. To show the performance of the proposed method, we increased the error of the pseudo measurements from 30\% to 50\% while keeping the same number of measurements. As shown in Fig. 9, the PAWNN and P2N2 show a better result when compared with the WLS. This means that the neural network model with a large set of training data can provide reliable estimation performance. The third scenario is carried out with a limited number of measurements, and 14 load power measurements are neglected (compared with the first scenario). In this case, the network is unobservable because of the limited number of measurements, and thus, the WLS cannot obtain estimates for the voltage magnitudes. However, the PAWNN and P2N2-based neural network show effectiveness even with the unobservable distribution system. Further, the average estimation errors of different scenarios are shown in Table I. It shows the accuracy of the WLS in the case of the observable distribution network with noiseless measurements. However, the better estimation result is achieved by PAWNN and P2N2 in the case of higher noise from the measurements or when the network is unobservable. Table II summarizes the estimation time of each time step, where the P2N2 method is nearly 160 times faster than the WLS.
\begin{table}[t]
\caption{The Average Estimation Accuracy}
\begin{center}
\begin{tabular}{|c|c|c|c|}
\hline
\textbf{Scenario}&\textbf{WLS}&\textbf{PAWNN}&\textbf{P2N2}\\
\hline
1&0.0019 & 0.0345 &0.0346\\
\hline
2&0.1188 & 0.0344 &0.0346\\
\hline
3&	-        &	0.0342 &0.0347\\
\hline
\end{tabular}
\label{tab2}
\vspace{-7mm}
\end{center}
\end{table}

\begin{table}[t]
\caption{The Estimation Time of Each Time Step}
\begin{center}
\begin{tabular}{|c|c|c|c|}
\hline
\textbf{Scenario}&\textbf{WLS}&\textbf{PAWNN}&\textbf{P2N2}\\
\hline
1&2.7299 s& 0.0198 s&0.0173 s\\
\hline
2&2.8212 s& 0.0189 s &0.0171 s\\
\hline
3&	-          & 0.0172 s &0.0156 s\\
\hline
\end{tabular}
\label{tab3}
\vspace{-7mm}
\end{center}
\end{table}

\section{Conclusions}
This paper proposed a data-driven state estimation method for the distribution system. The model was designed based on the physical connections of the distribution network, which pruned the unnecessary connections between layers. One-year smart meter data were used to generate the training data set by performing the power flow analysis. Then, the set of 35,040 data points were collected for the training and testing phase. Three different scenarios were carried out in the IEEE 123-node test network to show the performance of the proposed method. Numerical results show the efficacy of P2N2 in terms of reliable performance under different observability scenarios. Compared with WLS, the proposed method achieves better estimation accuracy in low-observability scenarios. Also, the proposed P2N2 approach can achieve almost the same performance as PAWNN while having a significant reduction in the number of parameters, thus reducing the training effort. Several extensions are possible to improve the method in the future. For example, system parameters can be exploited to design more efficient learning models.

\vspace{0mm}
\bibliographystyle{IEEEtran} 
\bibliography{Ref_PowerTech} \vspace{-2mm}

\end{document}